\pdfoutput=1
\documentclass[usenatbib]{mn2e}
\usepackage{siunitx}
\usepackage{graphicx}
\usepackage{latexsym}
\usepackage{aas_macros}
\usepackage{amsfonts,amsmath,amssymb}
\usepackage{url}
\usepackage[utf8]{inputenc}
\usepackage{fancyref}
\usepackage{subfloat}
\usepackage{verbatim}

\title[Modelling light curves in presence of flickering]{Using Gaussian processes to model light curves in the presence of flickering: the eclipsing cataclysmic variable ASASSN-14ag}

\author[M.\,J.\ McAllister et al.]{M.\,J.\ McAllister$^{1}$, S.\,P.\ Littlefair$^{1}$, V.\,S.\ Dhillon$^{1,2}$, T.\,R.\ Marsh$^{3}$, R.\,P.\ Ashley$^{3}$, 
\newauthor M.\,C.\,P.\ Bours$^{3,4}$, E.\ Breedt$^{3}$, L.\,K.\ Hardy$^{1}$, J.\,J.\ Hermes$^{5}$, S.\ Kengkriangkrai$^{6}$,
\newauthor P.\ Kerry$^{1}$, S.\ Rattanasoon$^{1,6}$, D.\,I.\ Sahman$^{1}$\\
$^{1}$Dept of Physics and Astronomy, University of Sheffield, Sheffield, S3 7RH, UK\\
$^{2}$Instituto de Astrofisica de Canarias, E-38205 La Laguna, Tenerife, Spain\\
$^{3}$Dept of Physics, University of Warwick, Coventry, CV4 7AL, UK\\
$^{4}$Departmento de F\'isica y Astronom\'ia, Universidad de Valpara\'iso, Avenida Gran Bretana 1111, Valpara\'iso, 2360102, Chile\\
$^{5}$Hubble Fellow - Department of Physics, University of North Carolina, Chapel Hill, NC 27599, USA\\
$^{6}$National Astronomical Research Institute of Thailand, 191 Siriphanich Building, Huay Kaew Road, Chiang Mai 50200, Thailand}

\begin{document}



\maketitle

\label{firstpage}

\begin{abstract}

The majority of cataclysmic variable (CV) stars contain a stochastic noise component in their light curves, commonly referred to as flickering. This can significantly affect the morphology of CV eclipses and increases the difficulty in obtaining accurate system parameters with reliable errors through eclipse modelling. Here we introduce a new approach to eclipse modelling, which models CV flickering with the help of Gaussian processes (GPs). A parameterised eclipse model - with an additional GP component - is simultaneously fit to 8 eclipses of the dwarf nova ASASSN-14ag and system parameters determined. We obtain a mass ratio $q$\,=\,0.149\,$\pm$\,0.016 and inclination $i$\,=\,83.4\,$^{+0.9}_{-0.6}$\,$^{\circ}$. The white dwarf and donor masses were found to be $M_{w}$\,=\,0.63\,$\pm$\,0.04\,$M_{\odot}$ and $M_{d}$\,=\,0.093\,$^{+0.015}_{-0.012}$\,$M_{\odot}$, respectively. A white dwarf temperature $T_{w}$\,=\,14000\,$^{+2200}_{-2000}$\,K and distance $d$\,=\,146\,$^{+24}_{-20}$\,pc were determined through multicolour photometry. We find GPs to be an effective way of modelling flickering in CV light curves and plan to use this new eclipse modelling approach going forward.

\end{abstract}


\bibliographystyle{mn2e_fixed2}

\section{Introduction} 
\label{sec:introduction}

Cataclysmic variable stars (CVs) are interacting binary systems that contain a white dwarf primary and a low mass secondary. Material from the secondary star is transferred to the white dwarf due to the secondary filling its Roche lobe. If the white dwarf has a low magnetic field, this transferred mass does not immediately accrete onto the white dwarf. Instead, in order to conserve angular momentum, the transferred mass forms an accretion disc around the white dwarf. A bright spot is formed at the point on the accretion disc where the gas stream from the donor makes contact. For a general review of CVs, see \cite{hellier01}.

At high enough inclinations to our line of sight (\textgreater\,80$^{\circ}$), the donor star can eclipse all other components within the system. As this includes the white dwarf, accretion disc and bright spot, CV eclipses can appear complex in shape. All of these components are eclipsed in quick succession, therefore high-time resolution photometry is required to reveal all the individual eclipse features. Measuring the timings of the white dwarf and bright spot eclipse features allow the system parameters to be accurately determined (e.g. \citealt{wood86}).

For some systems, the timing of these features (especially those associated with the bright spot) cannot be accurately measured, even with high-time resolution. This can be due to such systems containing a high amount of flickering, seen as random variability in CV light curves with amplitudes reaching the same order of magnitude as the bright spot eclipse features. Flickering in CVs is found to originate in both the bright spot and the inner accretion disc, and is due to the turbulent nature of the transferred material within the system \citep{bruch00,bruch15,baptistabortoletto04,scaringi12,scaringi14}.

Previous photometric studies of eclipsing CVs have used the averaging of multiple eclipses as a way of overcoming flickering and strengthening the bright spot eclipse features, before fitting an eclipse model to obtain system parameters (e.g. \citealt{savoury11,littlefair14,mcallister15}). \cite{mcallister15} also attempted to estimate the effect of flickering on the parameter uncertainties. An additional four $g'$-band eclipses were created -- each containing a different combination of three out of the four original eclipses used for the $g'$-band average -- and fit separately. The spread in system parameters from these average eclipses gave an indication of the error due to flickering, approximately five times the size of the purely statistical error.
 
 A downside to the eclipse averaging approach concerns the inconsistent bright spot ingress/egress positions due to changes in the accretion disc radius, which are observed in a significant number of systems. Averaging such light curves can lead to inaccurate bright spot eclipse timings and therefore incorrect system parameters. Eclipse light curves from systems with disc radius changes have to be fit individually, requiring another method to combat flickering. Here we introduce a new approach, involving the modelling of flickering in individual eclipses with the help of Gaussian processes (GPs). 

GPs have been used for many years in the machine learning community (see textbooks: \citealt{rasmussenwilliams06,bishop06}), and have recently started seeing use in many areas of astrophysics. Some examples include photometric redshift prediction \citep{waysrivastava06,way09}, modelling instrumental systematics in transmission spectroscopy \citep{gibson12,evans15} and modelling stellar activity signals in radial velocity studies \citep{rajpaul15}. See section~\ref{sec:gps} for further discussion of GPs.

The modelling of flickering is just one of a number of modifications we have made to the fitting approach. The model now has the ability to fit multiple eclipses simultaneously, whilst sharing parameters intrinsic to a particular system, e.g mass ratio ($q$), white dwarf eclipse phase full-width at half-depth ($\Delta\phi$) and white dwarf radius ($R_{w}$) between all eclipses. More details on the modifications to the model can be found in section~\ref{subsec:modifications}.

ASASSN-14ag was the chosen system to test the new modelling approach, due to the combination of a high level of flickering and clear bright spot features in its eclipse light curves. ASASSN-14ag was discovered in outburst (reaching $V$=13.5) by the All-Sky Automated Search for Supernovae (ASAS-SN; \citealt{shappee14}) on 14th March 2014. A look through existing light curve data on this system from the Catalina Real-Time Survey (CRTS; \citealt{drake09}) showed signs of eclipses, with an orbital period below the period gap (vsnet-alert 17036). Follow up photometry made in the days following the initial ASAS-SN discovery confirmed the eclipsing nature of the CV (vsnet-alert 17041). The discovery of superhumps also showed this to be a superoutburst, identifying ASASSN-14ag as a SU UMa-type dwarf nova (vsnet-alert 17042; \citealt{kato15}).

\section{Observations}
\label{sec:observations}

ASASSN-14ag was observed a total of 14 times from Nov 2014 -- Dec 2015 using the high-speed single beam camera ULTRASPEC \citep{dhillon14} on the 2.4-m Thai National Telescope (TNT), Thailand. Eclipses were observed in the SDSS $u' g' r' i'$ and Schott KG5 filters. The Schott KG5 filter is a broad filter, covering approximately $u' + g' + r'$. A complete journal of observations is shown in Table~\ref{table:obs}.

Data reduction was carried out using the ULTRACAM pipeline reduction software (see \citealt{feline04}). A nearby, bright and photometrically stable comparison star was used to correct for any transparency variations during observations.

The standard stars SA 92-288 (observed on 1st Jan 2015), SA 97-249 (2nd Jan 2015), SA 93-333 (3rd Jan 2015 \& 11th Dec 2015), SA  97-351 (3rd Mar 2015) and SA 100-280 (10th Dec 2015) were used to transform the photometry into the $u' g' r' i' z'$ standard system \citep{smith02}. The KG5 filter was calibrated using a similar method to \cite{bell12}; see appendix of Hardy et al. (2016, submitted) for a full description of the calibration process. A KG5 magnitude was calculated for the SDSS standard star SA 97-249 (27 Feb 2015), and used to find a target flux in the KG5 band. Photometry was corrected for atmospheric extinction using extinction values -- for all bands -- measured at the observatory \citep{dhillon14}.

\begin{table*}
\begin{center}
\begin{tabular}{lcccccccccc}
\hline
Date & Start Phase & End Phase & Cycle & Filter & $T_{mid}$ & $T_{exp}$ & $N_{exp}$ & Seeing & Airmass & Phot?\\
&&& No. && (HMJD) & (seconds) && (arcsecs) && \\ \hline
2014 Nov 27 & -35.304 & -34.854 & -35 & KG5 & 56988.75612(3) & 1.964 & 1186 & 1.8-2.1 & 1.48-1.80 & Yes \\
2014 Nov 29 & -0.107 & 0.321 & 0 & KG5 & 56990.86702(3) & 1.964 & 1124 & 1.0-1.4 & 1.06-1.07 & No \\
2014 Nov 30 & 15.793 & 16.244 & 16 & KG5 & 56991.83195(3) & 1.964 & 1188 & 1.3-2.5 & 1.09-1.14 & No \\
2015 Jan 01 & 544.755 & 545.201 & 545 & $g'$ & 57023.73631(4) & 1.964 & 1177 & 1.2-2.1 & 1.11-1.18 & Yes \\
2015 Jan 02 & 559.867 & 560.202 & 560 & $r'$ & 57024.64101(4) & 1.964 & 883 & 1.2-2.0 & 1.64-1.96 & Yes \\ 
2015 Jan 03 & 579.878 & 580.177 & 580 & $i'$ & 57025.84724(4) & 1.964 & 789 & 0.9-1.2 & 1.12-1.17 & Yes \\
2015 Jan 04 & 593.865 & 594.267 & 594 & $g'$ & 57026.69153(4) & 1.964 & 1061 & 1.5-2.3 & 1.20-1.33 & No \\
2015 Jan 04 & 596.866 & 597.163 & 597 & $r'$ & 57026.87251(4) & 2.964 & 521 & 1.1-1.7 & 1.21-1.30 & Yes \\
2015 Feb 24 & 1440.645 & 1441.253 & 1441 & $g'$ & 57077.77465(3) & 3.964 & 795 & 1.6-3.0 & 1.36-1.72 & No \\
2015 Feb 25 & 1454.707 & 1455.215 & 1455 & $r'$ & 57078.61896(3) & 3.352 & 787 & 1.2-2.0 & 1.06-1.10 & No \\
2015 Feb 26 & 1473.891 & 1474.343 & 1474 & $g'$ & 57079.76494(3) & 3.964 & 594 & 1.6-2.7 & 1.44-1.74 & Yes \\
2015 Mar 03 & 1554.742 & 1555.271 & 1555 & $i'$ & 57084.64993(10) & 4.852 & 569 & 1.2-2.3 & 1.06-1.10 & No \\
2015 Dec 05 & 6149.849 & 6150.155 & 6150 & $u'$ & 57361.77768(8) & 9.564 & 169 & 2.1-2.9 & 1.21-1.30 & No \\
2015 Dec 07 & 6182.701 & 6183.148 & 6183 & $u'$ & 57363.76780(8) & 9.564 & 246 & 2.0-2.8 & 1.23-1.40 & No \\
\hline
\end{tabular}
\caption{\label{table:obs}Journal of observations. The dead-time between exposures was 0.015\,s for all observations. The relative timestamping accuracy is of order 10\,$\mu$s, while the absolute GPS timestamp on each data point is accurate to $<$\,1\,ms. $T_{mid}$ represents the mid-eclipse time, while $T_{exp}$ \& $N_{exp}$ represent the exposure time and number of exposures, respectively. The last column indicates whether or not the conditions were photometric.}
\end{center}
\end{table*}

\section{Gaussian processes}
\label{sec:gps}

Our aim here is to only briefly cover the topic of GPs, as they are covered extensively elsewhere in the literature. We recommend the textbooks of \cite{rasmussenwilliams06} and \cite{bishop06} as general overviews of the topic, while useful introductions to the use of GPs for modelling time-series data can be found in \cite{roberts13} and the appendix of \cite{gibson12}.

In the same way that a single datapoint can be represented by a Gaussian random variable, a light-curve of observables $\textbf{y}$ can be represented by a multivariate Gaussian distribution, which is completely specified by the mean values, $\boldsymbol\mu$, and a covariance matrix, $\textbf{K}$. Trends in the light curve are captured by correlations between nearby data points; i.e off-diagonal entries in the covariance matrix. The covariance matrix is represented by:

\begin{equation}
\textbf{K}_{ij} = \sigma^{2}_{i}\delta_{ij} + k(t_{i}, t_{j})
\label{eq:covmatrix}
\end{equation}
consisting of a white noise component, $\sigma^{2}_{i}\delta_{ij}$, and a covariance function, $k(t_{i}, t_{j})$. The covariance function determines the covariance between any two data points, and is chosen to best represent the stochastic process to be modelled. For modelling flickering in CV light curves, the Mat\'ern-3/2 kernel was favoured over the more commonly used squared-exponential kernel. This is due to the Mat\'ern-3/2's greater ability at recreating the sharp features of flickering that comes from being finitely differentiable. The Mat\'ern-3/2 kernel has the following form:

\begin{equation}
k(t_{i}, t_{j}) = h^{2}\left(1+\sqrt3 \frac{|t_{i}-t_{j}|}{\lambda}\right) {\rm exp} \left(-\sqrt 3 \frac{|t_{i}-t_{j}|}{\lambda}\right)	  	
\label{eq:maternkernel}
\end{equation}
where $k(t_{i}, t_{j})$ is the $ij^{th}$ element of $k$ and $t_{i}$ and $t_{j}$ represent the times of any two data points \citep{roberts13}. Both $h$ and $\lambda$ are $hyperparameters$ of the GP, and they control the output scale (amplitude) and input scale (time), respectively. Once a kernel function has been constructed, it is straightforward to calculate the likelihood, $\mathcal{L}$, of a dataset:

\begin{equation}
\log\mathcal{L}_{\mu,k}(h,\lambda) = -\frac{1}{2}\textbf{r}^{\rm{T}}\textbf{K}^{-1}\textbf{r}\,-\,\frac{1}{2}\log|\textbf{K}|\,-\,\frac{n}{2}\log(2\pi)
\label{eq:likelihood}
\end{equation}
where $\textbf{r} = \textbf{y} - \boldsymbol\mu$ represents the vector of the residuals after subtraction of the mean function, $\boldsymbol\mu$, from the data, \textbf{y}, and $n$ is the number of data points (\citealt{rasmussenwilliams06}). The mean and uncertainty of the GP can also be calculated given observed data, i.e. the posterior mean and uncertainty (see equations 8 \& 9 in \citealt{roberts13}). Equation~\ref{eq:likelihood} is expensive to compute due to the need for inverting the $N \times N$ covariance matrix, requiring $\mathcal{O}(n^{3})$ operations. For large matrices, it is possible to speed up this step by using an alternative solver based on an $\mathcal{O}(n\log^{2}n)$ algorithm for inversion \citep{ambikasaran14}

As mentioned in section~\ref{sec:introduction}, there are multiple sources of flickering in CVs, and therefore more than one flickering amplitude. The observed amplitude should vary across the eclipse as the different components are individually eclipsed. GPs are stationary, and therefore act the same across all points in the time-series. To accommodate for the anticipated changes in flickering amplitude, two $changepoints$ were introduced. These changepoints are positioned at the white dwarf's ingress start, $t_{in}$, and egress end, $t_{e}$. This enabled the kernel function amplitude hyperparameter outside white dwarf eclipse, $h_{1}$, to differ from that inside, $h_{2}$. The location of the changepoints was chosen on the basis that the inner disc is a main source of flickering, but not the only source. The input scale hyperparameter was kept the same across the whole time-series. The drastic changepoint approach from \cite{garnett10} was implemented, with the kernel function taking the following form:

\small
\begin{equation}
k(t_{1},t_{2};h_{1},h_{2}) \triangleq \left\{
	\begin{array}{lr}
	k(t_{1},t_{2};h_{1}), & t_{1},t_{2} < t_{in}\\
	k(t_{1},t_{2};h_{2}), & t_{1},t_{2} \ge t_{in} ; t_{1},t_{2} \le t_{e}\\
	k(t_{1},t_{2};h_{1}), & t_{1},t_{2} > t_{e} \\
	0, & \rm otherwise.
	\end{array}
\right.
\label{eq:changepoints}
\end{equation}
\normalsize

\section{Results}
\label{sec:results}

\subsection{Orbital ephemeris}
\label{subsec:orbeph}

Mid-eclipse times ($T_{mid}$) were determined assuming that the white dwarf eclipse is symmetric around phase zero: $T_{mid} = (T_{wi} + T_{we})/2$, where $T_{wi}$ and $T_{we}$ are the times of white dwarf mid-ingress and mid-egress, respectively. $T_{wi}$ and $T_{we}$ were determined by locating the minimum and maximum times of the smoothed light curve derivative. The $T_{mid}$ errors (see Table~\ref{table:obs}) were adjusted to give $\chi^{2}$ = 1 with respect to a linear fit.

All eclipses were used to determine the following ephemeris:
\\
\\
$HMJD$ = 56990.867004(12) + 0.060310665(9) $E$
\\
\\
This ephemeris was used to phase-fold the data for the analysis that follows.

\subsection{Light curve morphology and variations}
\label{subsec:lcmorph} 

\begin{figure}
\begin{center}
\includegraphics[width=1.0\columnwidth,trim=10 10 10 10]{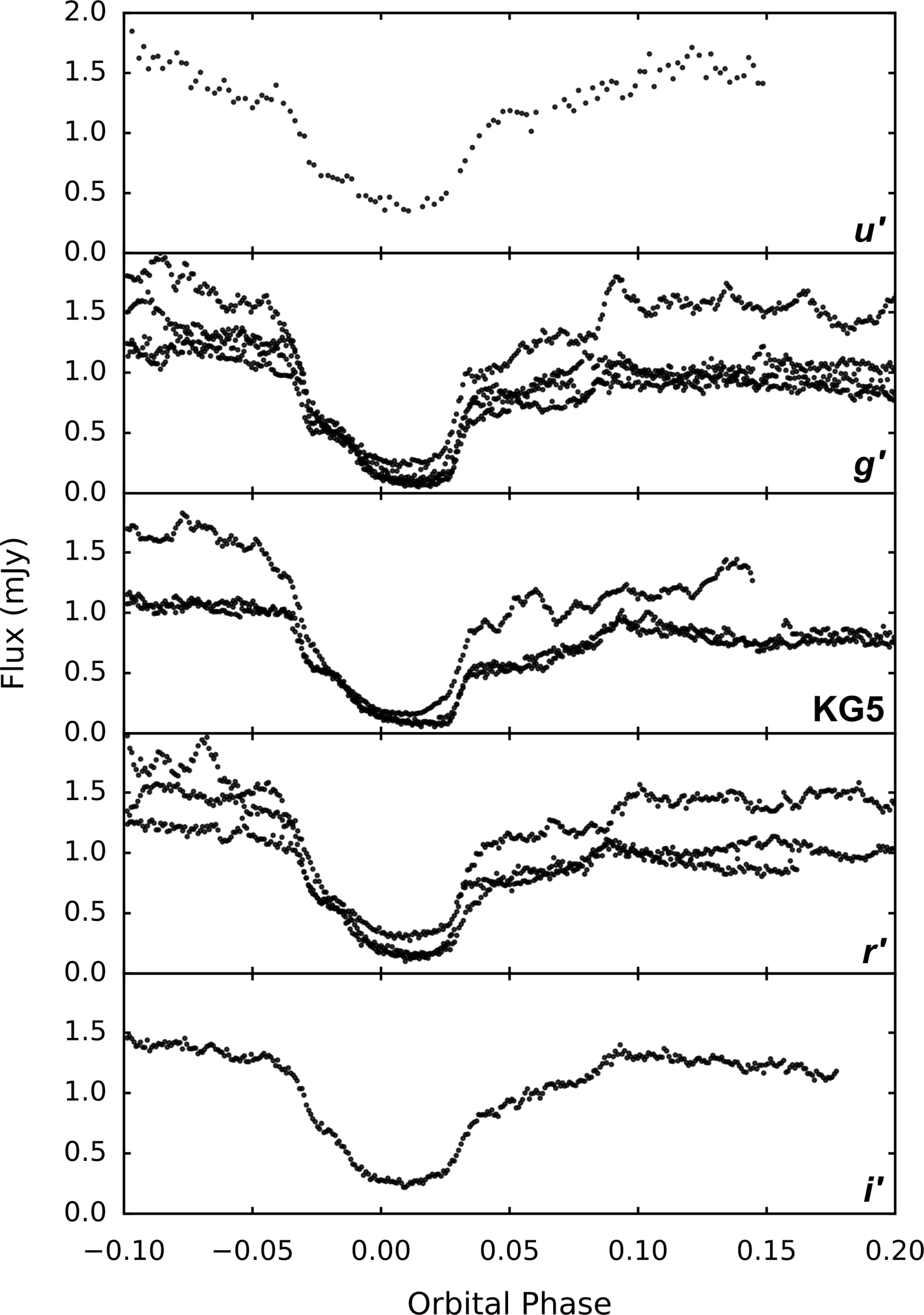}
\caption{\label{fig:ecl}All 12 ASASSN-14ag eclipses that are not badly affected by atmospheric conditions. Each plot contains those eclipses observed in one of the 5 wavelength bands, with the name of each band in the bottom-right corner of each plot.}
\end{center}
\end{figure}

Figure~\ref{fig:ecl} shows 12 of the 14 total ASASSN-14ag eclipses. The eclipses of 03 Mar 2015 and 05 Dec 2015 were affected by poor atmospheric conditions, so were not used in this study. The eclipses in Figure~\ref{fig:ecl} all have a clear white dwarf eclipse feature (phase -0.03 to 0.03), and the majority also have a discernible bright spot eclipse feature (phase -0.02 to 0.08). The positions of bright spot ingress and egress appear to occur at slightly different phases in each eclipse. This may be evidence for small changes in the accretion disc radius or could be due to flickering, which is  inherent to every eclipse and of varying amplitude from one eclipse to the next.

The majority of the flickering occurs outside of white dwarf eclipse. In some cases it re-appears almost immediately after white dwarf egress, implying the source of flickering to be in proximity to the white dwarf, perhaps in either the inner disc or the boundary layer. In a number of eclipses there is a small amount of flickering visible between the two ingress features. As the white dwarf is eclipsed during this period, there must be another source of flickering within the system. Flickering is greatly reduce once both the white dwarf and bright spot are eclipsed, which points to the bright spot as the secondary source of flickering.

The highest amplitude flickering is seen in the three eclipses that were observed while the system was in a slightly higher photometric state, with one such eclipse in each of the KG5, $g'$ and $r'$ bands (Figure~\ref{fig:ecl}). The higher photometric state is most likely to be the result of a more luminous disc. The high state $g'$ and $r'$ band eclipses do show a clear bright spot egress feature, but an ingress is not visible in any of the three eclipses and therefore none were included for model fitting.

\subsection{Modifications to existing model}
\label{subsec:modifications}

The model of the binary system used to calculate eclipse light curves contains contributions from the white dwarf, bright spot, accretion disc \& secondary star, and is described in detail by \cite{savoury11}. The model requires a number of assumptions, including that of an unobscured white dwarf \citep{savoury11}. As stated in \cite{mcallister15}, we feel this is still a reasonable assumption to make, despite the validity of the assumption being questioned by \cite{sparkodonoghue15} through fast photometry observations of the dwarf nova OY Car.

We have made modifications to this model so that it is now possible to fit multiple eclipse light curves simultaneously, with the $q$, $R_{w}/a$ and $\Delta\phi$ parameters shared between all eclipses. Each eclipse in the simultaneous fit also has either 11 or 15 (depending on whether the simple or complex bright spot model is used; \citealt{savoury11}) parameters that are unique to that eclipse. Due to the prominence of the bright spot in ASASSN-14ag, the complex bright spot model was used in all fits. The three shared parameters, once constrained through model fitting, can then be used to calculate system parameters (see section~\ref{subsubsec:syspars}).

With GPs included to model the flickering, the total number of model parameters are increased by three with the inclusion of the three kernel function hyperparameters (see section~\ref{sec:gps}). When fitting the eclipse model, the flickering is handled by using the residuals -- obtained by subtracting the eclipse model from the data -- to calculate the model likelihood using equation~\ref{eq:likelihood}.

\subsection{Simultaneous light curve modelling}
\label{subsec:simlcmod}

\begin{figure*}
\begin{center}
\includegraphics[width=1.8\columnwidth,trim=20 10 20 20]{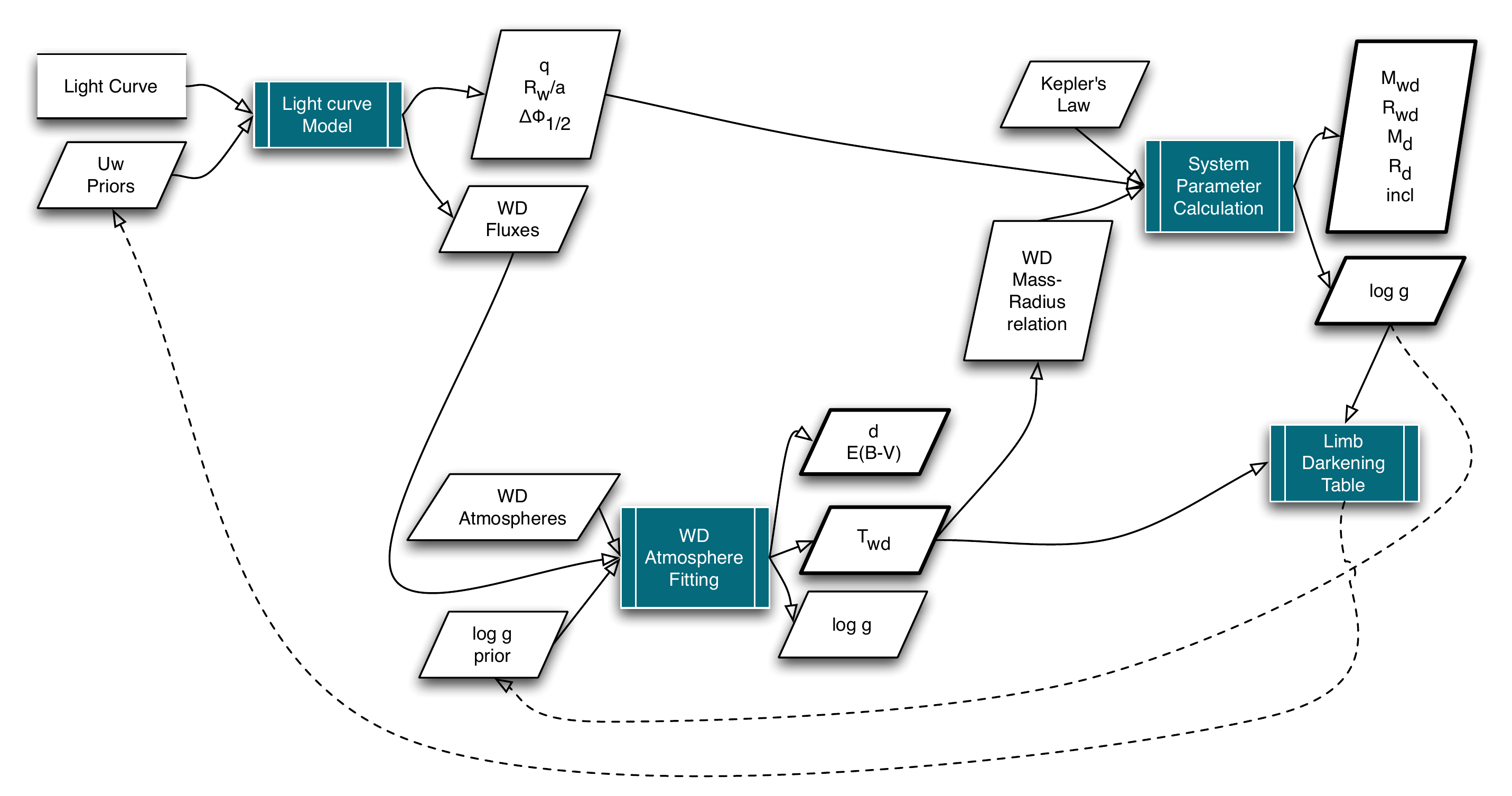}
\caption{\label{fig:flowchart} A schematic of the eclipse fitting procedure used to obtain system parameters. Two iterations of the fitting procedure occur, the dotted lines show steps to be taken only during the first iteration.}
\end{center}
\end{figure*}

Discarding the two eclipses affected by poor atmospheric conditions and the three eclipses in the higher photometric state left a total of 9 eclipses to use for modelling. The 8 of these eclipses taken in bands other than $u'$ were simultaneously fit with the model, both with and without the use of GPs. The $u'$ band eclipse was not used in the simultaneous fit as a consequence of its lower signal-to-noise and time-resolution compared to other wavelength bands, although it was fit separately (see below). All 123 parameters (126 in the GP case) were left to fit freely, except for the 8 limb-darkening parameters ($U_{w}$). This is due to our data not being of sufficient cadence and signal-to-noise to enable the shape of the white dwarf ingress/egress features to be determined; a requirement for $U_{w}$ to be accurately constrained. The $U_{w}$ parameter's priors were heavily constrained around values determined from a preliminary run through of the fitting procedure described below and shown schematically in Figure~\ref{fig:flowchart}.

A parallel-tempered Markov chain Monte Carlo (MCMC) ensemble sampler \citep{earldeem05,foreman-mackey13} was used to draw samples from the posterior probability distribution of the model parameters. A parallel-tempered sampler was chosen due to the large number of model parameters to be fit, and therefore large size of parameter space. Parallel-tempering involves multiple MCMCs running simultaneously, all at different `temperatures', $T$. Each MCMC samples from a modified posterior:

\begin{equation}
\pi_{T}(x) = [l(x)]^{\frac{1}{T}}p(x)
\label{eq:ptsampler}
\end{equation}
where $l(x)$ and $p(x)$ represent the likelihood and prior functions, respectively. As equation~\ref{eq:ptsampler} shows, each MCMC's likelihood function scales to the power of the temperature's reciprocal, so chains at higher temperatures can explore parameter space much more effectively. Communication between each MCMC occurs through chains at adjacent temperatures periodically swapping members of their ensemble \citep{earldeem05}. This greatly assists convergence to a global solution. A total of 10 MCMCs -- the first of temperature one and all others a factor of $\sqrt{2}$ higher than the one before -- were ran for 7500 steps. The first 5000 of these steps took the form of a burn-in phase and were discarded. Only the MCMC with a temperature equal to one at the end of the fit was used to produce the model parameter posterior probability distributions. The Gelman-Rubin diagnostic was used to confirm convergence \citep{gelmanrubin92}.

\subsubsection{White dwarf atmosphere fitting}
\label{subsubsec:wdatmos}

Estimates of the white dwarf temperature, log\,$g$ and distance were obtained through fitting white dwarf fluxes -- at $u'$, $g'$, $r'$, $i'$ and KG5 wavelengths -- to white dwarf atmosphere predictions \citep{bergeron95} with an affine-invariant MCMC ensemble sampler \citep{goodmanweare10,foreman-mackey13}. Reddening was also included as a parameter, in order for its uncertainty to be taken into account, but is not constrained by our data. Its prior covered the range from 0 to the maximum galactic extinction along the line-of-sight \citep{schlaflyfinkbeiner11}. The $g'$, $r'$, $i'$ \& KG5 white dwarf fluxes and errors were taken as median values and standard deviations from a random sample of the simultaneous 8-eclipse fit chain. The $u'$ band flux was obtained through an individual fit to the 7th Dec 2016 $u'$ band eclipse, keeping $q$, $R_{w}/a$ and $\Delta\phi$ parameters close to their values from the simultaneous fit with Gaussian priors. A 3\% systematic error was added to the fluxes to account for uncertainties in photometric calibration.

Knowledge of the white dwarf temperature and log\,$g$ values enabled the estimation of the $U_{w}$ parameters, with use of the data tables in \cite{gianninas13}. Linear limb-darkening parameters of 0.427, 0.369, 0.317 and 0.272 were determined for $u'$, $g'$, $r'$ and $i'$ bands, respectively. A value of 0.360 for the KG5 band was calculated by taking a weighted mean of the $u'$, $g'$ and $r'$ values, based on the fraction of the KG5 bandpass covered by each of the three SDSS filters.

\subsubsection{System parameters}
\label{subsubsec:syspars}

The posterior probability distributions of $q$, $\Delta\phi$ and $R_{w}/a$ returned by the MCMC eclipse fit described in section~\ref{subsec:simlcmod} were used along with Kepler's third law, the system's orbital period and a temperature-corrected white dwarf mass-radius relationship \citep{wood95}, to calculate the posterior probability distributions of the system parameters \citep{savoury11}, which include:

\begin{enumerate}
\item mass ratio, $q$;
\item white dwarf mass, $M_{w}$;
\item white dwarf radius, $R_{w}$;
\item white dwarf log\,$g$;
\item donor mass, $M_{d}$;
\item donor radius, $R_{d}$;
\item binary separation, $a$;
\item white dwarf radial velocity, $K_{w}$;
\item donor radial velocity, $K_{d}$;
\item inclination, $i$.
\end{enumerate}

The most likely value of each distribution is taken as the value of each system parameter, with upper and lower bounds derived from 67\% confidence levels.

The system parameters were calculated twice in total. The value for log\,$g$ returned from the first calculation was used to constrain the log\,$g$ prior in a second MCMC fitting the white dwarf fluxes to model atmosphere predictions \citep{bergeron95}, as described in section~\ref{subsubsec:wdatmos}. All of these steps are shown schematically in Figure~\ref{fig:flowchart}. Constraining log\,$g$ had little effect on the the white dwarf temperature in this instance, although it significantly helped the distance estimate, which was very poorly constrained after the first fit.

\subsubsection{Without GPs}
\label{subsubsec:nogps}

\begin{subfigures}
\begin{figure*}
\begin{center}
\includegraphics[width=2.0\columnwidth,trim=80 60 0 80]{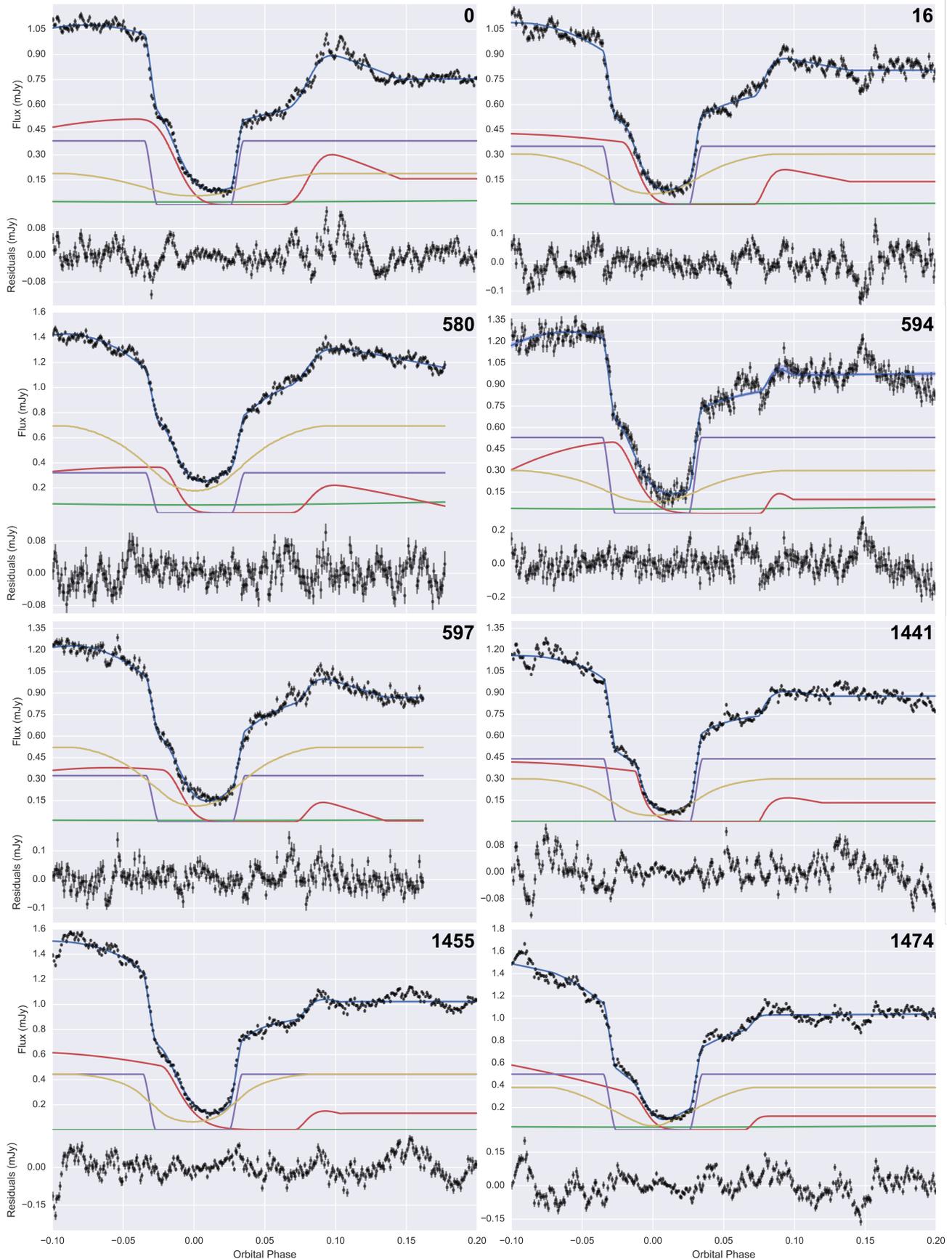}
\caption{\label{fig:allecl_nogps} Simultaneous model fit (blue) to 8 ASASSN-14ag eclipses (black) without GPs. The blue fill-between region represents 1$\sigma$ from the mean of a random sample (size 1000) of the MCMC chain, although this is thinner than the model fit line in all but one case (cycle no. 594). Also shown are the different components to the model: white dwarf (purple), bright spot (red), accretion disc (yellow) and donor (green). The residuals are shown at the bottom of each plot. Cycle numbers are displayed at the top-right corner of each plot.}
\end{center}
\end{figure*}

\begin{figure*}
\begin{center}
\includegraphics[width=2.0\columnwidth,trim=80 60 0 80]{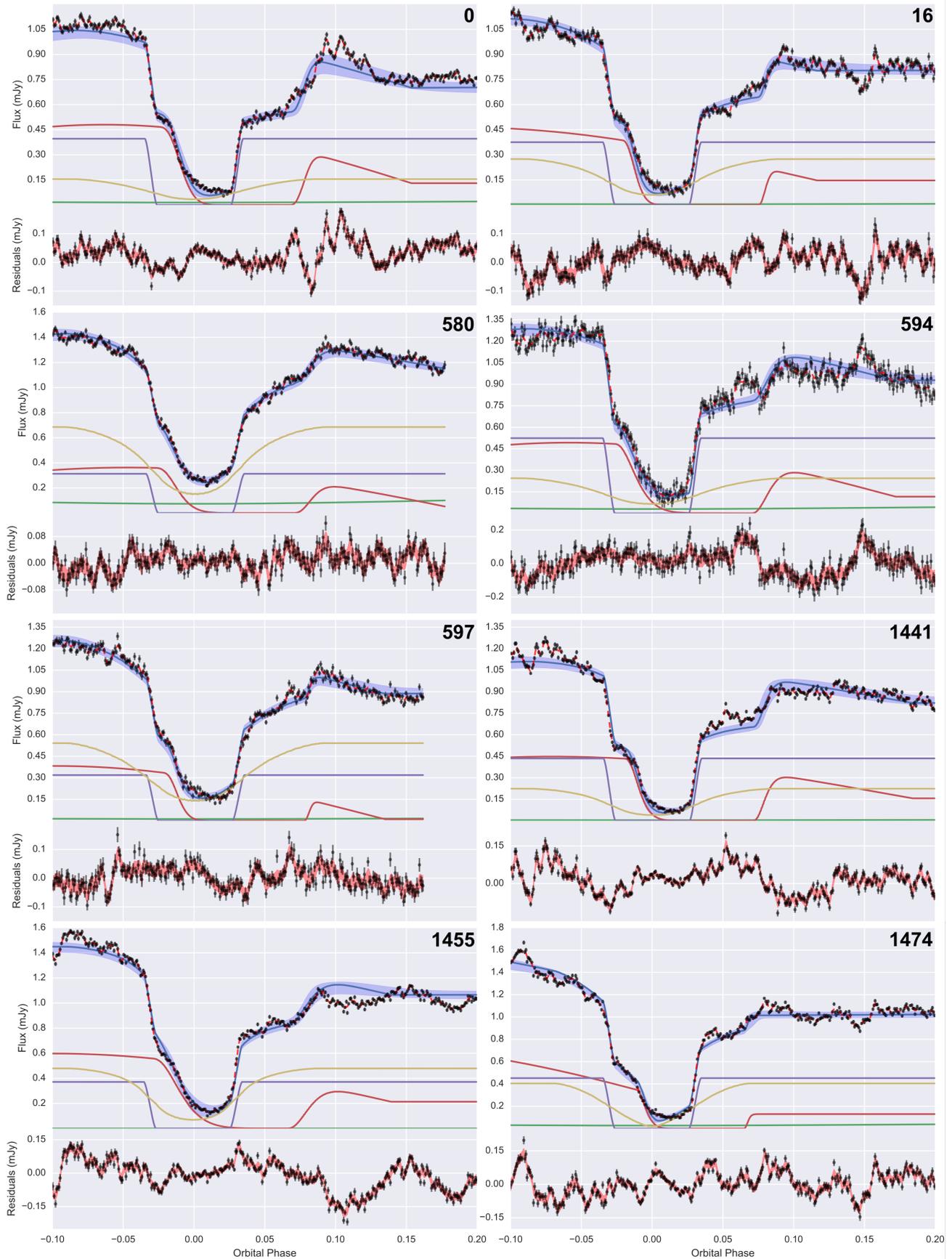}
\caption{\label{fig:allecl_gps} Simultaneous model fit (blue) to 8 ASASSN-14ag eclipses (black) with GPs included. The blue fill-between region represents 1$\sigma$ from the mean of a random sample (size 1000) of the MCMC chain. The red dashed line shows the sum of the eclipse model +  posterior mean of the GP. Also shown are the different components to the model: white dwarf (purple), bright spot (red), accretion disc (yellow) and donor (green). The residuals are shown at the bottom of each plot, with the red fill-between region covering 2$\sigma$ from the posterior mean of the GP. Cycle numbers are displayed at the top-right corner of each plot.}
\end{center}
\end{figure*}
\end{subfigures}

\begin{figure*}
\begin{center}
\includegraphics[width=2.0\columnwidth,trim=20 10 20 10]{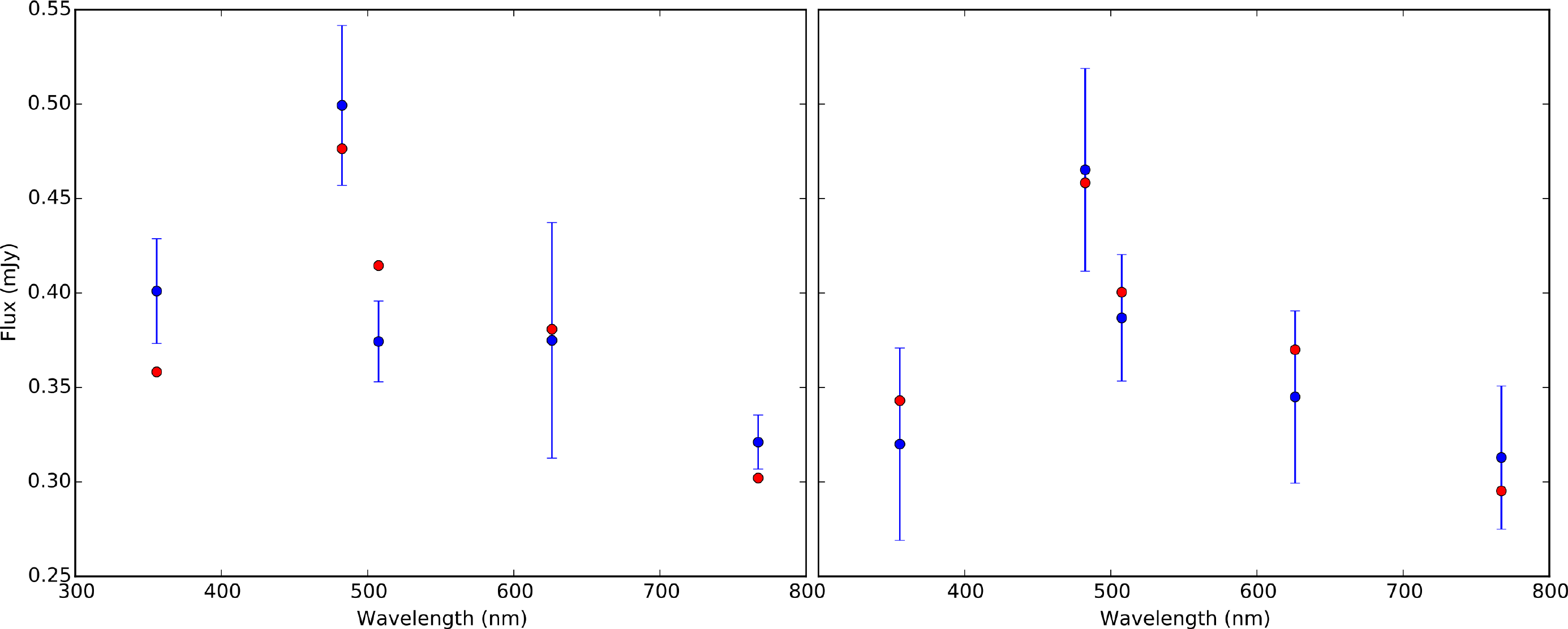}
\caption{\label{fig:fluxes} White dwarf fluxes from both simultaneous 8-eclipse \& individual $u'$-band model fits (blue) and \protect\cite{bergeron95} white dwarf atmosphere predictions (red), at wavelengths corresponding to (from left to right) $u'$, $g'$, KG5, $r'$ \& $i'$ filters. The left plot is  without GPs and the right plot is with GPs.}
\end{center}
\end{figure*}

\begin{figure}
\begin{center}
\includegraphics[width=1.0\columnwidth,trim=10 10 10 10]{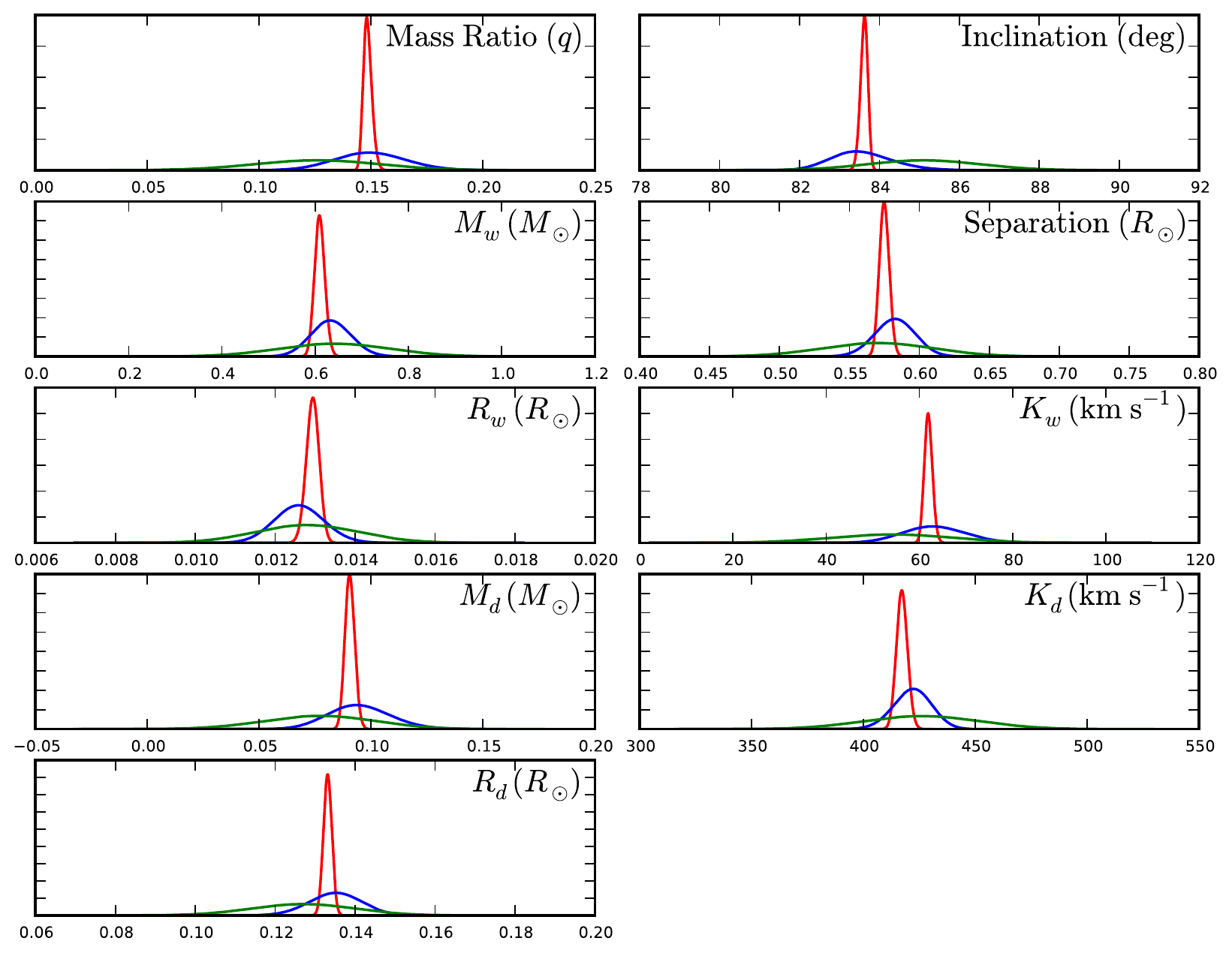}
\caption{\label{fig:pdf} Normalised posterior probability density functions for both simultaneous 8-eclipse fits: without GPs (red) and with GPs (blue). Also included (green) are the parameter distributions calculated from the average eclipse fits (see section~\ref{subsec:avlcmod}).}
\end{center}
\end{figure}

The model fit to 8 eclipses simultaneously, without the use of GPs is shown in Figure~\ref{fig:allecl_nogps}. The most probable fit to the 8 eclipses has a $\chi^{2}$ of 17363 with 2598 degrees of freedom. This large value of $\chi^{2}$ is due to the large amount of flickering present in each eclipse. The model appears to fit every white dwarf eclipse reasonably well, although there are one or two cases (e.g. cycle nos. 1441 \& 1474) where the depth of the white dwarf eclipse has been overestimated slightly. In general, the bright spot eclipses have also been fit fairly well. The major exception to this is cycle no. 0. This eclipse contains low amplitude flickering until a slight brightening prior to bright spot egress and then a significant flicker just afterwards covering the orbital phases 0.08-0.13. These two features, especially the large flicker, make it impossible for the model to correctly fit the bright spot egress. This flicker is modelled as part of the eclipse structure, at the expense of bright spot ingress, which is free from flickering but poorly fit to accommodate for egress. Another notable bright spot ingress mis-fitting is in cycle no. 1455. The ingress feature is not as clear as in other eclipses, and the nature of the flickering before the eclipse results in a falsely high bright spot flux, hindering bright spot ingress fitting. Overall, the model fits to every eclipse have been affected by flickering to some degree.

Also plotted in Figure~\ref{fig:allecl_nogps} is a fill-between region representing 1$\sigma$ from the mean of a random sample (size 1000) of the MCMC chain. In all but one case, this fill-between region is not visible due to it being thinner than the blue line of the most probable model fit. The exception to this is cycle no. 594, where it is visible just after bright spot egress (phase\,$\sim$\,0.08). The very small distribution from a sample in the chain indicates a precise solution, and this is reflected in the very small errors associated with the model parameters returned by the fit.

A comparison between the measured white dwarf fluxes and models can be found in the left-hand plot of Figure~\ref{fig:fluxes}. The majority of the points calculated from the white dwarf atmosphere predictions lie outside the flux error bars, evidence for the underestimation of flux errors by the eclipse model due to flickering.

The posterior probability distributions for each system parameter are displayed in red in Figure~\ref{fig:pdf}, with their peak values and associated errors -- as well as temperature and distance estimates -- given in the second column of Table~\ref{table:syspars_sim}. The probability distributions are very narrow, which translate to small errors on the system parameters.

\subsubsection{With GPs}
\label{subsubsec:gps}
 
Another set of simultaneous model fits to the 8 eclipses -- this time with GPs included -- is shown in Figure~\ref{fig:allecl_gps}. The most probable fit has a much higher $\chi^{2}$ of 31812 (with 2595 degrees of freedom), reflected in the greater amplitude residuals in Figure~\ref{fig:allecl_gps}. This is due to the additional GP component included in the fit, which models the residuals from the eclipse model fit but is not taken into account when calculating $\chi^{2}$. The GP is best visualised as the red fill-between region covering 2$\sigma$ from the GP's posterior mean that overlays the residuals below each eclipse in Figure~\ref{fig:allecl_gps}. The majority of residuals are covered by this fill-between region, which indicates that the chosen Mat\'ern-3/2 kernel provides a good description of CV flickering. Another representation of the GP takes the form of the sum of the eclipse model and mean of the GP, shown by a red dashed line on the main eclipse plots.

The most obvious difference with GPs is the significantly increased size of the blue fill-between region in each eclipse. As mentioned in section~\ref{subsubsec:nogps}, this represents 1$\sigma$ from the mean of a random sample (size 1000) of the MCMC chain. The increase in $\sigma$ is due to the much broader model parameter solution distributions, which is a consequence of fitting the model in accordance with GPs. A wider range of parameters are allowed by the data, as differences between the eclipse model and the data can be accommodated by the GP.

In contrast to the standard model fitting, it is the fill-between region, not the most probable fit, that is of most importance. In most cases, this region is in agreement with both the white dwarf and bright spot eclipse features. Cycle no. 0 is a good example to show what the GPs have brought to the model fitting. With just the eclipse model, the large flicker after bright spot egress created a very extended bright spot that couldn't successfully fit bright spot ingress (Figure~\ref{fig:allecl_nogps}). When GPs are used, the flicker can be modelled, allowing a much more compact bright spot and a correctly fit ingress.

As discussed in section~\ref{sec:gps}, our GP framework includes changepoints at the start and end of white dwarf eclipse, due to the expectation that the amplitude of flickering should differ inside and outside white dwarf eclipse. The GP amplitude inside white dwarf eclipse returned by the fit is an order of magnitude lower compared to that outside, validating the use of changepoints.

The white dwarf atmosphere prediction fit to white dwarf fluxes obtained using GPs is shown in the right-hand plot of Figure~\ref{fig:fluxes}. This is much improved compared to the previous fit (left-hand side of the same figure). Comparing the white dwarf fluxes from both the standard model and GP approaches, the only flux value to have changed significantly is that of the $u'$ band. A more realistic $u'$ band flux is obtained when fitting with GPs, as well as more representative errors in all bands.

The posterior probability distributions for each system parameter are shown in blue in Figure~\ref{fig:pdf}. The most likely parameter values and associated errors -- as well as temperature and distance estimates -- are shown in Table~\ref{table:syspars_sim}. It is very evident from Figure~\ref{fig:pdf} that, while the distributions from each fitting approach have similar peak values, they have very contrasting values of $\sigma$. This was already apparent from the differing sizes of the blue fill-between regions in Figures~\ref{fig:allecl_nogps} \&~\ref{fig:allecl_gps}, and results in the errors on the system parameters from the GP fit being significantly greater than those from the standard model fit.

\subsubsection{Accretion disc}
\label{subsubsec:disc}

\begin{figure*}
\begin{center}
\includegraphics[width=2.0\columnwidth,trim=20 10 20 0]{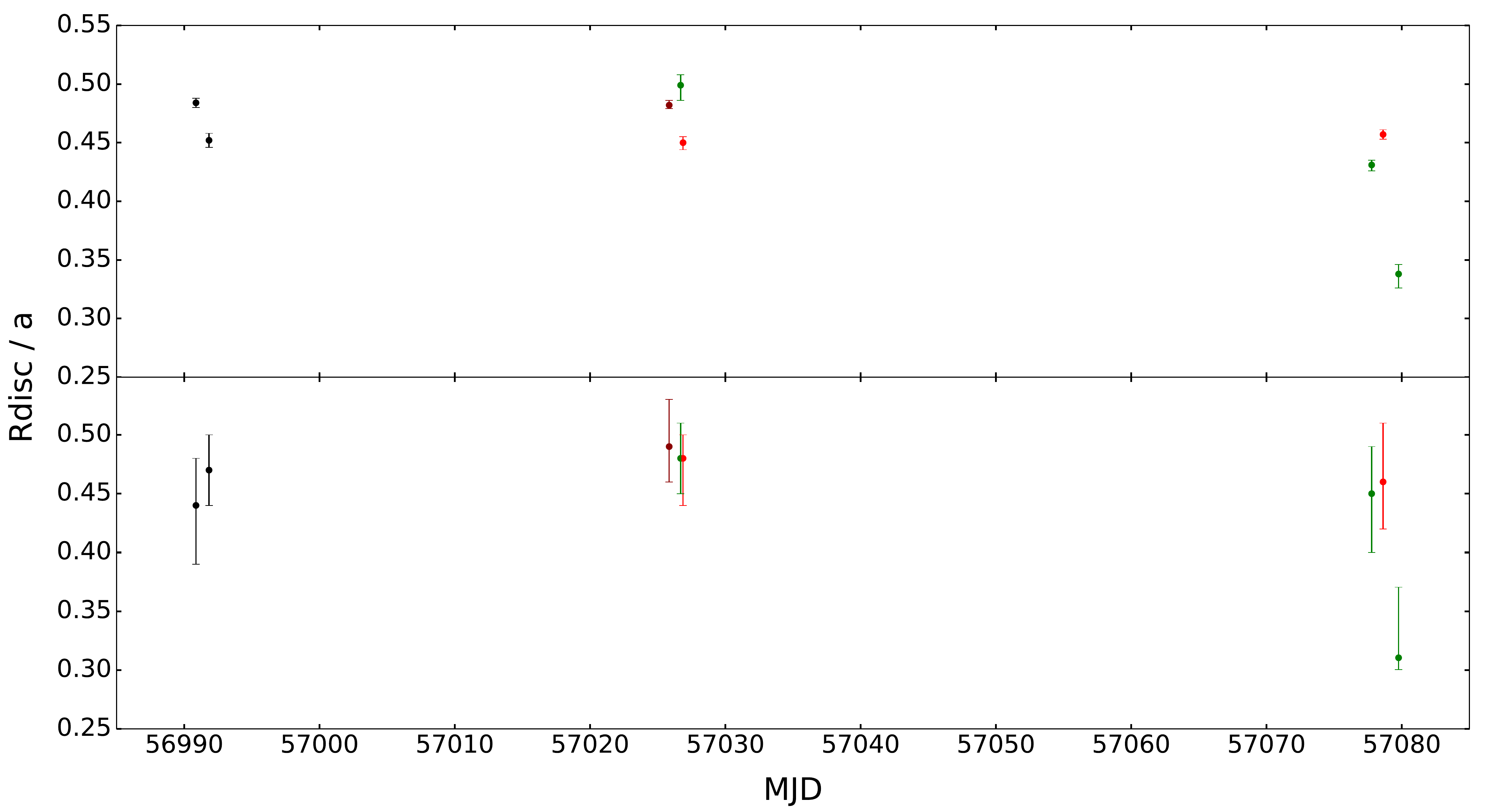}
\caption{\label{fig:discrad} Accretion disc radius ($R_{disc}$) as a fraction of the binary separation ($a$) vs time (in MJD) for the 8 eclipses fit simultaneously. The top panel shows radii from the non-GP fit, while the bottom panel shows radii from the GP fit. The colour of each data point shows the wavelength band its eclipse was observed in: KG5 (black), $g'$ (green), $r'$ (red), $i'$ (dark red).}
\end{center}
\end{figure*}

One of the parameters included in the model is the radius of the accretion disc as a fraction of the binary separation ($R_{disc}/a$). This value from the model is actually the bright spot's distance from the white dwarf as a fraction of the binary separation, but we assume the bright spot to lie at the edge of the accretion disc. Plotting these values from multiple eclipses against the MJD of each eclipse -- e.g. Figure 7 in \cite{mcallister15} -- enables disc radius evolution to be investigated. Figure~\ref{fig:discrad} shows $R_{disc}/a$ against MJD for the 8 eclipses fit simultaneously, both without (top) and with GPs (bottom). These eclipses were observed during the same observing season, but over three separate observing runs, explaining the large gaps in Figure~\ref{fig:discrad}. The errors on the disc radii from the GP case are much larger than those from the non-GP case, which is expected as the errors from the non-GP case are significantly underestimated due to not taking the effects of flickering into account. Across the first two observing runs, there's very little change in disc radius. This is still the case even when the first two eclipses from the third observing run are included, but not once the final of the 8 eclipses is considered. The top plot of Figure~\ref{fig:discrad} appears to show a significant decrease in disc radius of order 0.1\,$R_{disc}/a$ between the final two eclipses, separated by approximately 1 day (19 orbital cycles). With GPs included (bottom plot), this apparent decrease in disc radius is shown to most likely be just a product of flickering, with less than a 2$\sigma$ difference between the final two disc radii.

\subsection{Average light curve modelling}
\label{subsec:avlcmod}

\begin{figure}
\begin{center}
\includegraphics[width=1.0\columnwidth,trim=10 10 10 20]{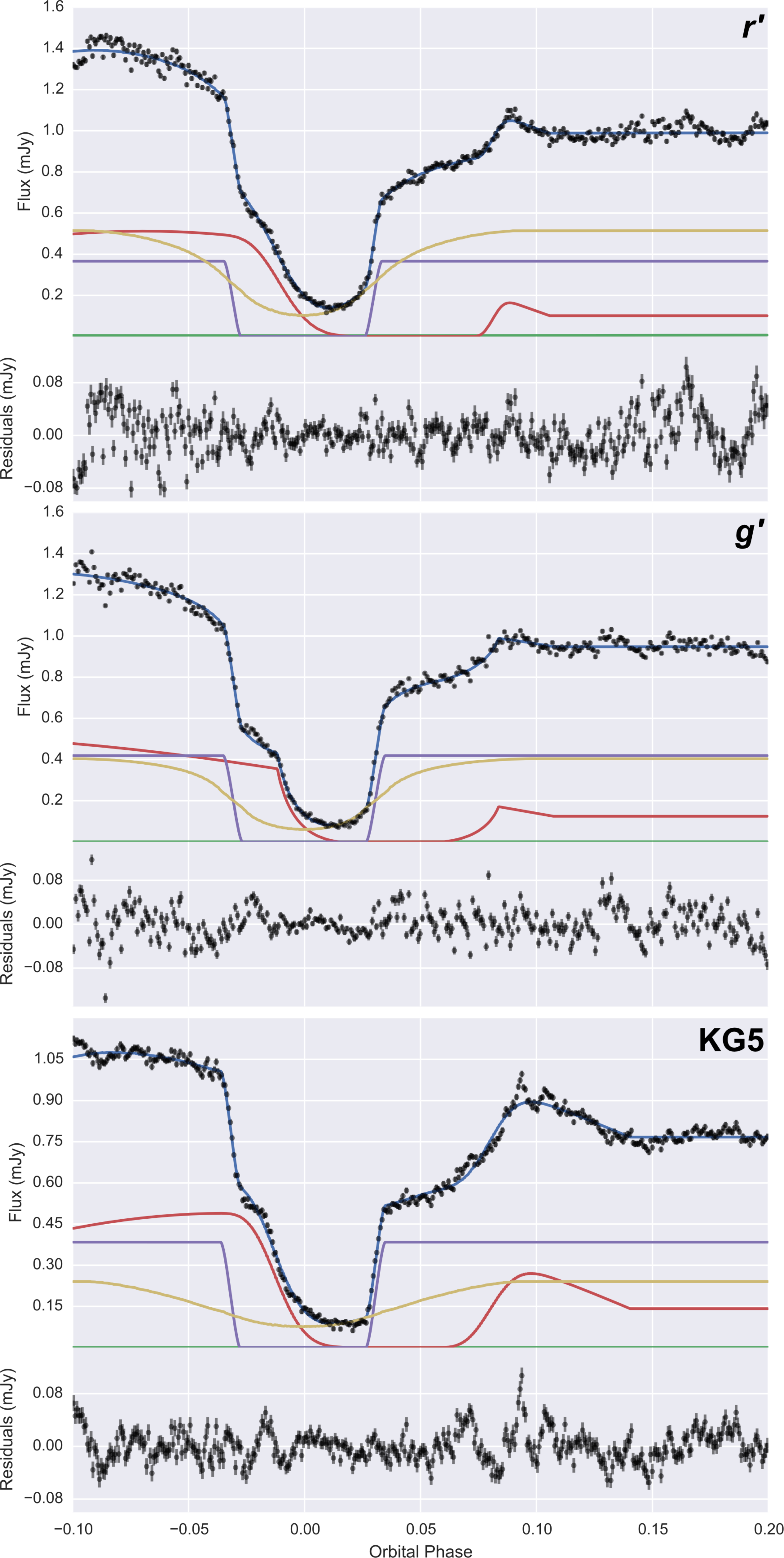}
\caption{\label{fig:av_fits} Model fits (blue) to average $r'$, $g'$ \& KG5 eclipses (black) without the use of GPs. Also shown are the different components to the model: white dwarf (purple), bright spot (red), accretion disc (yellow) and donor (green). The residuals are shown at the bottom of each plot. }
\end{center}
\end{figure}

Previous eclipse modelling studies have used the technique of averaging eclipses together as a way of negating flickering and/or boosting signal-to-noise \citep{littlefair08,savoury11,mcallister15}. To investigate how this technique compares to our new approach, the individual ASASSN-14ag eclipses were used to create average eclipses, which were then fit separately without the use of GPs. In addition, the 8 individual eclipses were also fit separately, with their spread in system parameters an indication of the effects of flickering.

The two KG5 band eclipses (cycle nos. 0 \& 16) were averaged to create a KG5 average eclipse, similarly for both the $g'$ (cycle nos. 594, 1441 \& 1474) and $r'$ (cycle nos. 597 \& 1455) bands (Figure~\ref{fig:av_fits}). Due to high-amplitude flickering and the small number of eclipses, averaging is relatively ineffective in this case, with all three average eclipses containing large amounts of residual flickering. Despite this, they do all show bright spot features. The three average eclipses were individually fit with the eclipse model (without GPs) and the resulting fits are shown in Figure~\ref{fig:av_fits}. The system parameters for each band can be found in the first three columns of Table~\ref{table:syspars_av}, with the final column in Table~\ref{table:syspars_av} showing the weighted mean of the parameters from each wavelength band. The errors on the parameter values in this final column are the weighted standard deviation of the parameters from fitting the 8 individual eclipses separately. The system parameter distributions from the average eclipse fitting are also shown in Figure~\ref{fig:pdf}. A similar -- though not identical -- approach to estimating the error due to flickering was used in \cite{mcallister15}.

\begingroup
\setlength{\tabcolsep}{10pt}
\renewcommand{\arraystretch}{1.5}
\begin{table*}
\begin{center}
\begin{tabular}{lcccc}
\hline
Parameter & Without GPs & With GPs \\ \hline
$q$ & 0.1480\,$^{+0.0023}_{-0.0014}$ & 0.149\,$\pm$\,0.016 \\
$M_{w}$ ($M_{\odot}$) & 0.609\,$^{+0.012}_{-0.010}$ & 0.63\,$\pm$\,0.04 \\
$R_{w}$ ($R_{\odot}$) & 0.01294\,$^{+0.00015}_{-0.00017}$ & 0.0126\,$\pm$\,0.0006 \\
$M_{d}$ ($M_{\odot}$) & 0.0903\,$^{+0.0023}_{-0.0020}$ & 0.093\,$^{+0.015}_{-0.012}$ \\
$R_{d}$ ($R_{\odot}$) & 0.1331\,$\pm$\,0.0011 & 0.135\,$\pm$\,0.007 \\
$a$ ($R_{\odot}$) & 0.575\,$\pm$\,0.004 & 0.583\,$\pm$\,0.015\\
$K_{w}$ (km\,s$^{-1}$) & 61.9\,$\pm$\,0.9 & 63\,$\pm$\,7 \\
$K_{d}$ (km\,s$^{-1}$) & 417.1\,$^{+2.6}_{-2.3}$ & 422\,$\pm$\,9 \\
$i$ $(^{\circ})$ & 83.63\,$^{+0.08}_{-0.11}$ & 83.4\,$^{+0.9}_{-0.6}$ \\ 
log\,$g$ & 7.999\,$\pm$\,0.013 & 8.04\,$\pm$\,0.05 \\ \hline
$T_{w}$ (K) & 14400\,$^{+1200}_{-1300}$ & 14000\,$^{+2200}_{-2000}$ \\
$d$ (pc) & 150\,$^{+14}_{-12}$ & 146\,$^{+22}_{-18}$ \\
\hline
\end{tabular}
\caption{\label{table:syspars_sim}System parameters for ASASSN-14ag through simultaneous fitting of 8 individual eclipses, with and without the use of GPs. $T_{w}$ and $d$ represent the temperature and distance of the white dwarf, respectively.}
\end{center}
\end{table*}
\endgroup

\begingroup
\setlength{\tabcolsep}{10pt}
\renewcommand{\arraystretch}{1.5}
\begin{table*}
\begin{center}
\begin{tabular}{lcccc}
\hline
Parameter & $g'$ & $r'$ & KG5 & Combined \\ \hline
$q$ & 0.113\,$^{+0.005}_{-0.001}$ & 0.1231\,$^{+0.0026}_{-0.0013}$ & 0.1301\,$^{+0.0022}_{-0.0007}$ &  0.126\,$\pm$\,0.028 \\
$M_{w}$ ($M_{\odot}$) & 0.602\,$\pm$\,0.019 & 0.70\,$\pm$\,0.04 & 0.608\,$^{+0.009}_{-0.017}$ & 0.64\,$\pm$\,0.12 \\
$R_{w}$ ($R_{\odot}$) & 0.01306\,$\pm$\,0.00028 & 0.0117\,$\pm$\,0.0004 & 0.01296\,$^{+0.00028}_{-0.00010}$ & 0.0128\,$\pm$\,0.0013 \\
$M_{d}$ ($M_{\odot}$) & 0.068\,$^{+0.005}_{-0.001}$ & 0.086\,$\pm$\,0.005 & 0.0794\,$^{+0.0015}_{-0.0023}$ & 0.077\,$\pm$\,0.025 \\
$R_{d}$ ($R_{\odot}$) & 0.1213\,$^{+0.0028}_{-0.0011}$ & 0.1313\,$\pm$\,0.0025 & 0.1276\,$^{+0.0008}_{-0.0013}$ & 0.127\,$\pm$\,0.013 \\
$a$ ($R_{\odot}$) & 0.567\,$\pm$\,0.006 & 0.597\,$\pm$\,0.010 & 0.571\,$^{+0.003}_{-0.005}$ & 0.57\,$\pm$\,0.04 \\
$K_{w}$ (km\,s$^{-1}$) & 47.8\,$^{+2.5}_{-0.2}$ & 54.8\,$^{+1.6}_{-1.2}$ & 55.1\,$\pm$\,0.7 & 54\,$\pm$\,13 \\
$K_{d}$ (km\,s$^{-1}$) & 426\,$\pm$\,4 & 443\,$\pm$\,7 & 422\,$^{+2}_{-4}$ & 426\,$\pm$\,26 \\
$i$ $(^{\circ})$ & 86.0\,$^{+0.1}_{-4.0}$ & 84.95\,$^{+0.09}_{-0.19}$ & 85.17\,$^{+0.06}_{-0.13}$ & 85.1\,$\pm$\,1.4 \\
log\,$g$ & 7.986\,$\pm$\,0.023 & 8.14\,$\pm$\,0.04 & 7.997\,$^{+0.020}_{-0.014}$ & 8.01\,$\pm$\,0.19 \\ \hline
\end{tabular}
\caption{\label{table:syspars_av}System parameters for ASASSN-14ag from average light curve fitting. The parameters in the combined column are calculated from the weighted mean of the values in each of the three bands. The errors on these combined parameters come from the weighted standard deviation of the parameters from the 8 individual eclipse fits.}
\end{center}
\end{table*}
\endgroup

\section{Discussion}
\label{sec:discussion}

As mentioned in section~\ref{subsec:avlcmod}, averaging eclipses in this particular case is not an effective way of reducing flickering, as a large amount of it still remains. Obtaining many more eclipses would help, but negating the flickering would not be possible due to its high-amplitude nature. This issue - coupled with the fact that many systems show disc radius changes - shows the need for a different approach to modelling CV light curves containing flickering. The approach we present here involves including flickering in the model, while fitting individual eclipses simultaneously.

\subsection{Modelling of flickering}
\label{subsec:flickering}

In \cite{mcallister15}, the effects of flickering were estimated looking at the spread in system parameters after fitting a further four average eclipses, each containing a different combination of three out of the four original eclipses used for the $g'$-band average. Here we use the spread in system parameters from fitting the 8 individual eclipses separately as an estimation of flickering. The resulting system parameters are shown in Table~\ref{table:syspars_av}. These individual eclipse fits show a wide spread in system parameters, which results in large errors.

Using the new technique of modelling flickering with GPs, the error introduced by flickering no longer has to be estimated, as it is already included in the errors on the system parameters returned by the model. In the case of ASASSN-14ag, the contribution from flickering is seen through comparison of the errors in the two columns of Table~\ref{table:syspars_sim} and the difference in the red and blue distributions in Figure~\ref{fig:pdf}.

The size of the error introduced by flickering differs depending on whether the old or new approach is used (blue and green distributions in Figure~\ref{fig:pdf}), but which comes closest to representing the true effect? Does the old approach of using the distribution in system parameters from individual fits overestimate the error due to flickering, or does the new approach of modelling flickering underestimate it?

Figure~\ref{fig:allecl_gps} shows that the GPs have done a good job of modelling the flickering. Therefore, the error on the model parameters -- which are marginalised over the GP hyperparameters -- are likely to be accurate. This suggests that the old approach of using the standard deviation of system parameters from fitting individual eclipses may over-estimate the error due to flickering.

\subsection{Component masses}
\label{subsec:wdmass}

The calculated mass of the white dwarf in ASASSN-14ag from both simultaneous fits -- with and without GPs -- are consistent, although the errors from the GP fit are much more representative of the real uncertainty in the measurement so we adopt those system parameters. The white dwarf mass in ASASSN-14ag is 0.63\,$\pm$\,0.04\,$M_{\odot}$, which is at the lower end for white dwarfs in CVs. ASASSN-14ag joins fellow eclipsing CVs HT Cas \citep{horne91} and SDSS J115207.00+404947.8 \citep{savoury11} in having a white dwarf below 0.7\,$M_{\odot}$ and approaching the mean white dwarf field mass of 0.621\,$M_{\odot}$ \citep{tremblay16}. The corresponding donor mass of 0.093\,$^{+0.015}_{-0.012}$\,$M_{\odot}$ is consistent with the main sequence donor evolutionary track from \cite{knigge11}. These component masses give ASASSN-14ag a $\sim$95\% chance of lying within the dynamically stable region in Figure 2 of \citealt{schreiber16}, in agreement with their empirical consequential angular momentum loss (eCAML) model that appears to solve multiple issues with CV evolution.

The white dwarf temperature and mass were used to calculate a medium-term average mass transfer rate of $\dot{M}\,=\,1.9\,^{+2.6}_{-1.1}\,\times\,10^{-10}\,M_{\odot}\,yr^{-1}$ \citep{townsleybildsten03,townsleygansicke09}, while ASASSN-14ag's orbital period of 1.44\,h was used to determine a secular mass transfer rate of $\dot{M}\,\sim\,0.6\,\times\,10^{-10}\,M_{\odot}\,yr^{-1}$ \citep{knigge11}. While these two values for the mass transfer rate are consistent, the slightly higher medium-term mass transfer rate indicates that the white dwarf temperature of 14000\,$^{+2200}_{-2000}$\,K is marginally hotter than we would expect.

\section{Conclusions}
\label{sec:conclusions}

We have introduced a new approach to modelling CV eclipses that enables multiple eclipses to be fit simultaneously, with the option to model any inherent flickering with GPs. This no longer requires eclipses to be averaged together in order to overcome the presence of flickering, a technique employed  in previous studies and subject to issues caused by disc radius changes.

This new approach has been tested using 8 eclipses of the eclipsing CV ASASSN-14ag. These eclipses -- all including flickering -- were fit simultaneously with and without GPs. Although both fits return a similar solution, the errors associated with the GP fit are much more representative given the large amount of flickering present.

We have shown GPs to be an effective way of modelling flickering, and plan to use this new eclipse modelling approach on many more eclipsing CV systems going forward.

\section{Acknowledgements}
\label{sec:acknowledgements}

We thank the anonymous referee for their useful comments. MJM acknowledges the support of a UK Science and Technology Facilities Council (STFC) funded PhD. SPL and VSD are supported by STFC grant ST/J001589/1. EB and TRM are supported by the STFC in the form of a Consolidated Grant. VSD and TRM acknowledge the support of the Leverhulme Trust for the operation of ULTRASPEC at the Thai National Telescope. The results presented in this paper are based on observations made at the Thai National Observatory, operated by the National Astronomical Research Institute of Thailand. This research has made use of NASA's Astrophysics Data System Bibliographic Services.

\bibliography{gp_references}

\end{document}